\documentclass[twocolumn,showpacs,preprintnumbers,amsmath,amssymb]{revtex4}
\usepackage{mathrsfs}
\usepackage{bbm}

\usepackage{graphicx}
\usepackage{dcolumn}
\usepackage{bm}
\usepackage{CJK}
\begin{document}
\title{Phonon spectrum and bonding properties of Bi$_{2}$Se$_{3}$: Role of strong spin-orbit interaction}
\author{Bao-Tian Wang$^{1}$}
\author{Ping Zhang$^{2}$}
\thanks{Author to whom correspondence should be addressed. E-mail: zhang\_ping@iapcm.ac.cn}
\affiliation{$^{1}$Institute of Theoretical Physics and Department
of Physics, Shanxi University, Taiyuan 030006, People's Republic of
China\\$^{2}$LCP, Institute of Applied Physics and Computational
Mathematics, Beijing 100088, People's Republic of China}

\begin{abstract}
Phonon dispersions of one typical three-dimensional topological insulator Bi$_{2}$Se$_{3}$ have been studied within density functional
theory. The soft modes of two acoustic branches along the $Z$$-$$F$ and
$\Gamma$$-$$F$ directions within the pure local density approximation will transit to show imaginary frequency oscillating after including the spin-orbit interaction (SOI). Similar phenomenon has also been observed for Bi$_{2}$Te$_{3}$. Besides, we have found that the weak van der Waals forces between two Se1 layers in Bi$_{2}$Se$_{3}$ are strengthened by turning on the SOI.
\end{abstract}
\maketitle

Topological insulators (TIs) have an energy gap in their bulk but, due
to SOI, possess one or more robust metallic states on their edge or
surface protected by time-reversal symmetry. \cite{HsiehPRL} This
novel quantum state, which can show various topological quantum effects
combined with its potential applications involving future spintronic
devices, quantum computing, and photonics, has attracted great
attention in condensed-matter physics and materials science
recently. \cite{XLQi1,Moore,Hasan,XLQi2} After revealing the topological
nature of the surface states in the first three-dimensional (3D) TI:
Bi$_{1-x}$Sb$_{x}$ by angle-resolved photoemission spectroscopy
(ARPES), the second generation materials: Bi$_{2}$Se$_{3}$,
Bi$_{2}$Te$_{3}$, and Sb$_{2}$Te$_{3}$ were theoretically predicted
to be the simplest 3D TIs whose surface states consist of a single
Dirac cone at the Gamma point \cite{ZhangHJ}. The topological nature
of these material has already been established by recent
experiments. \cite{Xia,Hsieh,Chen}

\begin{figure}[ptb]
\begin{center}
\includegraphics[width=0.8\linewidth]{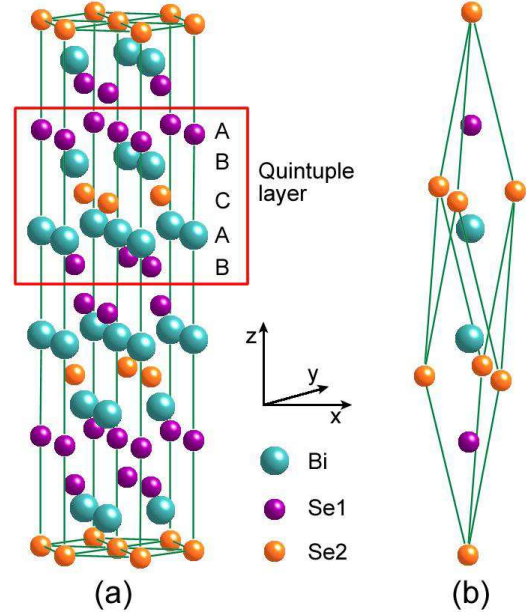}
\end{center}
\caption{(a) Hexagonal crystal structure of Bi$_{2}$Se$_{3}$ where a quintuple layer with Se1-Bi-Se2-Bi-Se1 sequence is presented by the red square. (b) Trigonal primitive cell of Bi$_{2}$Se$_{3}$.}%
\label{structure}%
\end{figure}

Here we focus our sight on Bi$_{2}$Se$_{3}$, in which an ARPES study
\cite{Xia} discovered a single surface electron pocket with a Dirac
point below the Fermi level. Actually, before the renewing interests due to its topological surface states, many efforts have been concentrated on this material owing to its
large thermoelectric effect. Early X-ray diffraction
experiment \cite{Nakajima} demonstrated that Bi$_{2}$Se$_{3}$
possesses layered structure along the hexagonal \emph{z} axis [see
Fig. 1(a)] with alternating layers of Bi and Se in Se1-Bi-Se2-Bi-Se1
order (known as quintuple layers). Transport and optical experiments
\cite{Black,Mooser,LaForge} have determined the insulting band gap
to be approximately 0.25-0.35 eV. Raman and far-infrared
\cite{Richter} as well as inelastic neutron scattering \cite{Rauh}
experiments have been conducted to investigate the phonon spectrum and phonon
density of state (DOS). Ultrafast carrier and phonon dynamics
\cite{Qi}, temperature dependent optical characters in a magnetic
field, \cite{LaForge} Faraday effect \cite{Sushkov}, and Raman response \cite{Gnezdilov} experiments of
Bi$_{2}$Se$_{3}$ have been studied in recent days. The easy cleavage between its neighboring quintuple
layers is due to the weak van der Waals forces between two Se1
layers. Strong covalent bonding character within each quintuple
layer has already been analyzed by performing density
functional theory (DFT) calculations. \cite{Mishra}
Our chemical bonding analysis will show that by including SOI the Se1-Se1 bonds can be viewed more strengthened than that described by pure DFT. The dynamical unstable feature of Bi$_{2}$Se$_{3}$ and Bi$_{2}$Te$_{3}$ is observed by investigating the phonon dispersions.

In this letter, the first-principles DFT calculations on the basis of the
frozen-core projected augmented wave (PAW) method of Bl\"{o}chl
\cite{PAW} are performed within the Vienna \textit{ab initio}
simulation package (VASP) \cite{Kresse3}, where the local density
approximation (LDA) \cite{LDA} and the Perdew, Burke, and Ernzerhof
(PBE) \cite{PBE} form of the generalized gradient approximation
(GGA) are employed to describe electron exchange and correlation. To
obtain accurate total energy, a cutoff energy of 300 eV is used for
the plane-wave set with a 10$\times$10$\times$10 \emph{k}
point-meshes in the Brillouin zone (BZ) for the
rhombohedral crystal structure. All atoms are fully
relaxed until the Hellmann-Feynman forces becoming less than 0.001
eV/\AA. The Bi 5$d$$^{10}$6$s$$^{2}$6$p$$^{3}$ and the Se
4$s$$^{2}$4$p$$^{4}$ orbitals are included as valence
electrons. SOI is included to compare with calculations without
considering SOI. Within LDA+SOI formalism, the symmetry of the
system is switched off.

\begin{table}[ptb]
\caption{Optimized structural parameters and insulating band gap
(\emph{E$_{g}$}) for Bi$_{2}$Se$_{3}$ without and with SOI. For
comparison, experimental values are also listed.}%
\begin{ruledtabular}
\begin{tabular}{lccccccccccccccc}
Method&\emph{a}&\emph{c}&\emph{u}&\emph{v}&E$_{g}$\\
&({\AA})&({\AA})&({\AA})&({\AA})&(eV)\\
\hline
LDA&4.102&27.865&0.4015&0.2090&0.0\\
LDA+SOI&4.112&27.611&0.4022&0.2074&0.3\\
Expt.&4.143$^{\emph{a}}$&28.636$^{\emph{a}}$&0.4008$^{\emph{a}}$&0.2117$^{\emph{a}}$&0.25$-$0.35$^{\emph{b}}$\\
\end{tabular}
$^{\emph{a}}$ Reference \cite{Nakajima}, $^{\emph{b}}$ References
\cite{Black,Mooser,LaForge} \label{lattice}
\end{ruledtabular}
\end{table}

Bi$_{2}$Se$_{3}$ crystallizes in rhombohedral crystal structure with
space group $R\bar{3}m$ (No. 166) with Bi in 2$c$($\mu$, $\mu$,
$\mu$), Se1 in 2$c$($\nu$, $\nu$, $\nu$), and Se2 in 1$a$(0, 0, 0)
Wyckoff positions. Using GGA and GGA+SOI, we obtain nonrational overestimations of the equilibrium volume and the band gap. But LDA
and LDA+SOI can give proper results of structural parameters in
comparison with experiments (see Table I) and also can successfully predict Bi$_{2}$Se$_{3}$ to be a topological insulator, as shown in
Fig. 2. Our calculated insulating band gap of about
0.3 eV within LDA+SOI formalism agrees well with recent
infrared reflectance and transmission measurements \cite{LaForge}
and previous calculations \cite{Mishra,ZhangHJ}.

\begin{figure}[ptb]
\begin{center}
\includegraphics[width=1.0\linewidth]{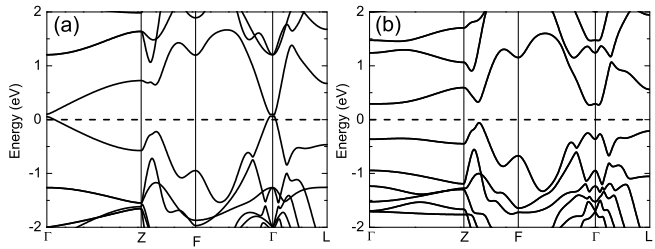}
\end{center}
\caption{Band structure for Bi$_{2}$Se$_{3}$ without (a) and with
(b) SOI. The Fermi energy level is set at
zero.}%
\label{bandstructure}%
\end{figure}

\begin{table*}[ptb]
\caption{Phonon frequencies at the $\Gamma$ point for Bi$_{2}$Se$_{3}$ and Bi$_{2}$Te$_{3}$ in unit of THz. For comparison, previous calculation results and experimental values are also listed.}%
\begin{ruledtabular}
\begin{tabular}{lccccccccccccccc}
&\multicolumn{4}{c}{Bi$_{2}$Se$_{3}$}&\multicolumn{4}{c}{Bi$_{2}$Te$_{3}$}\\
\cline{2-5}\cline{6-9}
Modes symmetry&LDA&LDA+SOI&LDA+SOI$^{\emph{a}}$&Expt.$^{\emph{b}}$&LDA&LDA+SOI&GGA+SOI$^{\emph{a}}$&Expt.$^{\emph{c}}$\\
\hline
$E_{g}^{\rm{I}}$&1.32&1.24&1.24&1.17&1.28&1.27&1.06&1.03\\
$A_{1g}^{\rm{I}}$&2.26&2.32&2.26&2.20&1.96&1.99&1.62&1.86\\
$E_{g}^{\rm{II}}$&4.31&4.15&4.11&3.98&2.38&3.20&2.88&3.05\\
$A_{1g}^{\rm{II}}$&5.45&5.27&5.13&5.26&4.26&4.04&3.81&4.02\\
$E_{u}^{\rm{I}}$&2.64&2.41&2.41&2.04&2.14&1.75&1.45&\\
$E_{u}^{\rm{II}}$&4.06&3.92&3.91&3.75&3.03&2.87&2.74&\\
$A_{2u}^{\rm{I}}$&4.31&4.16&4.12&3.87&3.15&2.82&2.85&\\
$A_{2u}^{\rm{II}}$&4.93&4.83&4.84&4.80&3.85&3.62&3.56&\\
\end{tabular}
$^{\emph{a}}$ Reference \cite{Cheng}, $^{\emph{b}}$ References
\cite{Gnezdilov}, $^{\emph{c}}$ References
\cite{Shahil} \label{lattice}
\end{ruledtabular}
\end{table*}

\begin{figure}[ptb]
\begin{center}
\includegraphics[width=1.0\linewidth]{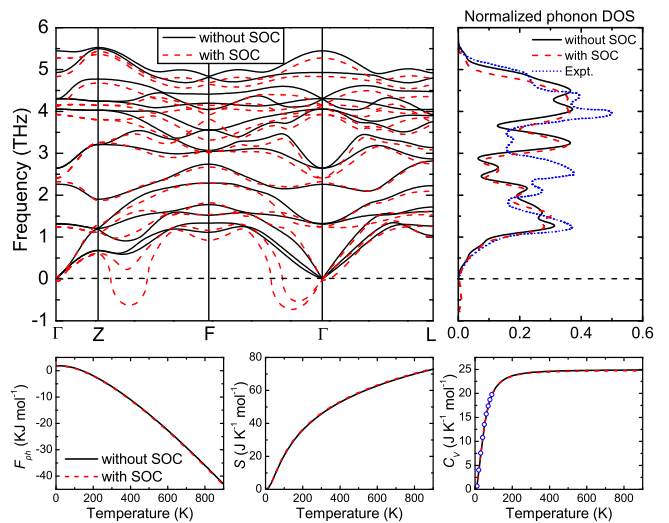}
\end{center}
\caption{Calculated phonon dispersion curves, phonon DOS, phonon
free energy ($F_{ph}$), entropy (\emph{S}), and specific heat at
constant volume ($C_{V}$) for Bi$_{2}$Se$_{3}$
without and with SOI. Experimental results of phonon DOS from Ref \cite{Rauh} and $C_{V}$ from Ref \cite{Mills} (hollow circles) are also presented.}%
\label{phononBi2Se3}%
\end{figure}

\begin{figure}[ptb]
\begin{center}
\includegraphics[width=1.0\linewidth]{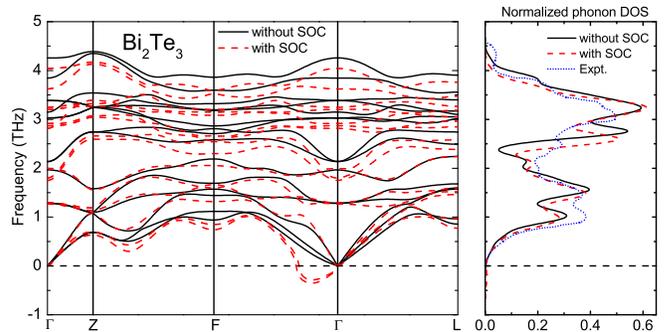}
\end{center}
\caption{Calculated phonon dispersion curves and phonon DOS for
Bi$_{2}$Te$_{3}$
without and with SOI. Experimental phonon DOS is taken from Ref \cite{Rauh}.}%
\label{phononBi2Te3}%
\end{figure}

Phonon spectrum has tight relation with dynamical stability, phase transition, thermoelectric effect, and superconductivity. In the present work, we compute the phonon spectrum for
Bi$_{2}$Se$_{3}$ with and without considering SOI. Phonon frequency
calculations are performed by using the supercell approach within the FROPHO code. \cite{Togo} To reach high accuracy,
we use 3$\times$3$\times$3 rhombohedral supercell containing 135
atoms and 2$\times$2$\times$2 Monkhorst-Pack \emph{k}-point mesh for
the BZ
integration. The calculated phonon curves along $\Gamma$%
$-$$Z$$-$$F$$-$$\Gamma$$-$$L$ directions and the phonon density of
states (DOS) are displayed in Fig. 3. For comparison, experimental
phonon DOS from Ref. \cite{Rauh} is also presented.

In the overall view of Fig. 3, most phonon curves at optical
branches calculated by LDA+SOI are a little lower than that without
including SOI. Two lowest acoustic branches exhibit soft oscillating
modes along the $Z$$-$$F$ and $\Gamma$$-$$F$ directions within LDA.
After including the SOI effect, this kind of soft modes transit to
show imaginary phonon frequencies, which illustrates dynamic
unstability of Bi$_{2}$Se$_{3}$ single crystal at low temperature.
This unstable nature has not been reported before, although some
experimental works \cite{Richter,Rauh,Qi,Gnezdilov} have focused on its lattice
vibration. To exclude the specificality of our results about
Bi$_{2}$Se$_{3}$, we have also calculated the phonon curves and
phonon DOS of Bi$_{2}$Te$_{3}$ (see Fig. 4). Clearly, the imaginary
phonon frequencies along $\Gamma$$-$$F$ direction after switching on
SOI can be seen. The good agreement between our calculated phonon
DOSs and the experimental results further guarantees the correctness
of our calculations. We notice that after structural analysis a
recent high-pressure work of Bi$_{2}$Te$_{3}$ \cite{Zhu} owed the
$\alpha$-Bi$_{2}$Te$_{3}$$\rightarrow$$\beta$-Bi$_{2}$Te$_{3}$ phase
transition mechanism to a parallel movement of BiTe$_{6}$ along the
[110] direction. Our phonon results support their conclusion and
supply direct vibration images for both Bi$_{2}$Se$_{3}$ and
Bi$_{2}$Te$_{3}$. The similarity of the phonon results for these two
typical 3D TIs suggests same phase transition path upon compression.
In Table II, we present the eigenfrequencies of
$A_{1g}^{\rm{I}}$, $A_{1g}^{\rm{II}}$, $E_{g}^{\rm{I}}$, $E_{g}^{\rm{II}}$, $E_{u}^{\rm{I}}$, $E_{u}^{\rm{II}}$, $A_{2u}^{\rm{I}}$, and $A_{2u}^{\rm{II}}$ phonon modes at Gamma point for Bi$_{2}$Se$_{3}$ and Bi$_{2}$Te$_{3}$ within LDA and LDA+SOI approaches. Good agreement with previous calculation \cite{Cheng} and experiments \cite{Richter,LaForge,Qi,Gnezdilov,Shahil} can be found. Compared with LDA, LDA+SOI improves the
agreement with experiments. As for the phonon DOS, relative strength
of three major peaks and two positions of them are successfully
reproduced in comparison with experiment. The second major peak is
relatively higher than experiment. These differences between our
calculations and experiments can be partially attributed to the
temperature effects. After all, the experiments were conducted under
300 and 77 K \cite{Richter,Rauh}. Besides, recent experiment
\cite{Qi} indicated that the air exposure can significantly affect
the carrier and phonon dynamics in Bi$_{2}$Se$_{3}$ crystals. The
temperature-induced electron-phonon interactions and defect-induced
charge trapping should be responsible for its dynamic stability.

Based on phonon DOS, we further obtain some thermodynamic properties
using the quasiharmonic approximation \cite{Siegel}, under which the
Helmholtz free energy \emph{F}(\emph{T,V}) at temperature \emph{T}
and volume \emph{V} can be expressed as
\begin{equation}
F(T,V)=E(V)+F_{ph}(T,V)+F_{el}(T,V),
\end{equation}
where \emph{E}(\emph{V}) is the ground-state total energy,
\emph{F$_{ph}$}(\emph{T,V}) is the phonon free energy and
\emph{F$_{el}$}(\emph{T,V}) is the thermal electronic contribution.
For present material, we focus only on the contribution of atom
vibrations. The \emph{F$_{ph}$}(\emph{T,V}) can be calculated by
\begin{equation}
F_{ph}(T,V)=k_{B}T\int_{0}^{\infty}g(\omega)\ln\left[  2 \sinh\left(
\frac{\hslash\omega}{2k_{B}T}\right)  \right]  d\omega,
\end{equation}
where $\hslash$ and $\emph{k}_{B}$ are Planck and Boltzmann
constants, respectively, $\omega$ represents the phonon frequencies
and $g(\omega)$ is the phonon DOS. Here, we only present results at
equilibrium volumes from both LDA and LDA+SOI. The entropy ($S$) can
be determined by $S$=$-T(\partial\ F/\partial\ T)_{V}$. The specific
heat at constant volume $C_{V}$ can be directly calculated through
\begin{equation}
C_{V}=k_{B}\int_{0}^{\infty}g(\omega)\left(  \frac{\hslash\omega}{k_{B}%
T}\right)  ^{2}\frac{\exp\frac{\hslash\omega}{k_{B}T}}{(\exp\frac
{\hslash\omega}{k_{B}T}-1)^{2}}d\omega.
\end{equation}
Calculated results of \emph{F$_{ph}$}(\emph{T}), $S$, and $C_{V}$
and previous experimental values of $C_{V}$ are shown in Fig. 3,
where it can be found that the calculated $C_{V}$ agrees well with
experiment \cite{Mills}. No evident differences have been observed between LDA
and LDA+SOI calculations. We hope that our results can provide
instruction for further study.

\begin{figure}[ptb]
\begin{center}
\includegraphics[width=1.0\linewidth]{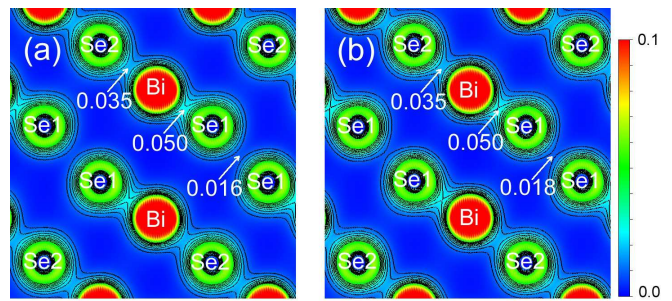}
\end{center}
\caption{Valence charge density of Bi$_{2}$Se$_{3}$ in the
\emph{xz}-plane without (a) and with SOI (b). Contour lines are
drawn from 0.0 to 0.1 at 0.01 e/{au}$^{3}$
intervals.}%
\label{charge}%
\end{figure}

\begin{table*}[ptb]
\caption{Calculated charge and volumes according to Bader
partitioning as well as the bond lengthes for Bi$_{2}$Se$_{3}$ without and with SOI.}%
\label{bader}
\begin{ruledtabular}
\begin{tabular}{cccccccccccccc}
Method&Q$_{B}$(Bi)&Q$_{B}$(Se1)&Q$_{B}$(Se2)&V$_{B}$(Bi)&V$_{B}$(Se1)&V$_{B}$(Se2)&Se1$-$Bi&Se2$-$Bi&Se1$-$Se1\\
&($e$)&($e$)&($e$)&({\AA}$^{3}$)&({\AA}$^{3}$)&({\AA}$^{3}$)&({\AA})&({\AA})&({\AA})\\
\hline
LDA&14.05&6.55&6.79&24.38&29.16&28.25&2.84&3.04&3.35\\
LDA+SOI&14.07&6.54&6.78&24.61&28.35&28.18&2.84&3.04&3.27\\
\end{tabular}
\end{ruledtabular}
\end{table*}

To understand the chemical-bonding characters of Bi$_{2}$Se$_{3}$,
we investigate the valence charge density distribution by plotting
plane charge density in the \emph{xz}-plane in both LDA and LDA+SOI
cases (see Fig. 5) as well as by performing the Bader analysis
\cite{Tang} (see Table III). The Se1$-$Bi, Se2$-$Bi, and
Se1$-$Se1 bond distances are also tabulated in Table III. The
correlated minimum values of charge density along the Se1$-$Bi,
Se2$-$Bi, and Se1$-$Se1 bonds are presented in Fig. 5. Obviously,
our present LDA results demonstrate that the binding between Bi and
Se layers exhibits predominant covalency and the interaction between
two Se1 layers is of the Van der Waals type, which coincides well
with previous conclusions. \cite{Rauh,Mishra} Comparing the results obtained by LDA and LDA+SOI, we find that the main influence on the structure, electronic distribution, and also the bonding nature of SOI lies between two Se1 layers. The weak van der Waals forces bonding of Se1-Se1 is strengthened by including SOI. The $z$ axis of the hexagonal crystal and the bond length of Se1-Se1 is shortened by SOI (see Tables I and III). As shown in Table III, the atomic volume of Se1 is diminished largely by LDA+SOI in comparison with the pure LDA calculations, while the increase or decrease amplitudes of V$_{B}$(Bi) and V$_{B}$(Se2) are not so evident. With SOI, the ionic charges of Bi$_{2}$Se$_{3}$ also
have a little change compared with LDA.
While Bi atom losses 0.95 electrons within LDA, it transfer 0.93
electrons to Se within LDA+SOI formalism. The ionic charges can be
represented as Bi$_{2}^{0.95+}$Se1$_{2}^{0.55-}$Se2$^{0.79-}$ and
Bi$_{2}^{0.93+}$Se1$_{2}^{0.54-}$Se2$^{0.78-}$ within LDA and
LDA+SOI, respectively. In two approaches, each Se2 atom gains more
electrons than Se1 atom.

In summary, we have demonstrated that the anti-crossing feature in the band structure around the $\Gamma$ point of Bi$_{2}$Se$_{3}$ can be successfully predicted by full geometry optimization within LDA and LDA+SOI formalisms. Good agreement of structural parameters and the eigenfrequencies of phonon modes at $\Gamma$ point for Bi$_{2}$Se$_{3}$ and Bi$_{2}$Te$_{3}$ between our theoretical results and experiments can be found. Within LDA+SOI, we have observed imaginary
phonon frequencies along $Z$$-$$F$ and $\Gamma$$-$$F$ directions for Bi$_{2}$Se$_{3}$ and along $\Gamma$$-$$F$ direction for Bi$_{2}$Te$_{3}$, which supports recent high-pressure work \cite{Zhu}. Their dynamic stability in experiments can be regarded as due to the temperature-induced electron-phonon interactions and defect-induced
charge trapping. We have found that the SOI plays an important role in describing the weak van der Waals forces between two Se1 layers.

This work was supported by
NSFC under Grant Nos. 11104170, 51071032, and 11074155, the
Foundations for Development of Science and Technology of China
Academy of Engineering Physics under Grant No. 2009B0301037.

\end{document}